\documentclass[9pt,twocolumns,letter]{elsart5p}
\pdfoutput=1
\usepackage{graphicx}
\usepackage{dcolumn}
\usepackage{bm}
\usepackage{hyperref}
\usepackage{color}
\usepackage{verbatim}

\newcommand{\ar}{\mbox{$^{39\!}$Ar}}
\newcommand{\arforty}{\mbox{$^{40\!}$Ar}}
\newcommand{\iso}[2]{\mbox{$^{#1}$#2}}

\begin{document}
\begin{frontmatter}

\title{Discovery of underground argon with low level of radioactive \ar\ \\
and possible applications to WIMP dark matter detectors}

\author[Princeton]{D.~Acosta-Kane}
\author[LAquila]{R.~Acciarri}
\author[Princeton]{O.~Amaize}
\author[LAquila]{M.~Antonello}
\author[Padova]{B.~Baibussinov}
\author[Padova]{M.~Baldo~Ceolin}
\author[Manchester]{C.\,J.~Ballentine}
\author[AirSep]{R.~Bansal}
\author[Otis]{L.~Basgall}
\author[Schlumberger]{A.~Bazarko}
\author[Pavia]{P.~Benetti}
\author[Princeton-ChemEng]{J.~Benziger}
\author[Princeton]{A.~Burgers}
\author[Princeton]{F.~Calaprice}
\author[Pavia]{E.~Calligarich}
\author[Pavia]{M.~Cambiaghi}
\author[LAquila]{N.~Canci}
\author[Napoli]{F.~Carbonara}
\author[Houston]{M.~Cassidy}
\author[LAquila]{F.~Cavanna}
\author[Padova]{S.~Centro}
\author[Princeton]{A.~Chavarria}
\author[Princeton]{D.~Cheng}
\author[Napoli]{A.\,G.~Cocco}
\author[NotreDame]{P.~Collon}
\author[Princeton]{F.~Dalnoki-Veress}
\author[Princeton]{E.~de~Haas}
\author[LAquila]{F.~Di~Pompeo}
\author[Napoli]{G.~Fiorillo}
\author[LindeEng]{F.~Fitch}
\author[Napoli]{V.~Gallo}
\author[Princeton]{C.~Galbiati\corauthref{cor}}
\corauth[cor]{Corresponding Authors}
\ead{galbiati@Princeton.EDU}
\author[Princeton]{M.~Gaull}
\author[LNGS]{S.~Gazzana}
\author[LNGS]{L.~Grandi}
\author[Princeton]{A.~Goretti}
\author[Otis]{R.~Highfill}
\author[Otis]{T.~Highfill}
\author[Princeton]{T.~Hohman}
\author[LNGS]{Al.~Ianni}
\author[Princeton]{An.~Ianni}
\author[Felician]{A.~LaCava}
\author[LNGS]{M.~Laubenstein}
\author[ANL]{H.\,Y.~Lee}
\author[Princeton]{M.~Leung}
\author[Princeton]{B.~Loer}
\author[Bern]{H.\,H.~Loosli}
\author[Princeton]{B.~Lyons}
\author[Princeton]{D.~Marks}
\author[Princeton]{K.~McCarty}
\author[Padova]{G.~Meng}
\author[Pavia]{C.~Montanari}
\author[Harvard]{S.~Mukhopadhyay}
\author[Princeton]{A.~Nelson}
\author[LNGS]{O.~Palamara}
\author[LNGS]{L.~Pandola}
\author[Padova]{F.~Pietropaolo}
\author[Otis]{T.~Pivonka}
\author[Stanford]{A.~Pocar}
\author[Bern]{R.~Purtschert\corauthref{cor}}
\ead{purtschert@climate.unibe.ch}
\author[Pavia]{A.~Rappoldi}
\author[Pavia]{G.~Raselli}
\author[Bicocca]{F.~Resnati}
\author[NotreDame]{D.~Robertson}
\author[Pavia]{M.~Roncadelli}
\author[Pavia]{M.~Rossella}
\author[LNGS]{C.~Rubbia}
\author[Princeton]{J.~Ruderman}
\author[Princeton]{R.~Saldanha}
\author[NotreDame]{C.~Schmitt}
\author[ANL]{R.~Scott}
\author[LNGS]{E.~Segreto}
\author[LindeGas]{A.~Shirley}
\author[Krakow,LAquila]{A.\,M.~Szelc}
\author[LNGS]{R.~Tartaglia}
\author[Princeton]{T.~Tesileanu}
\author[Padova]{S.~Ventura}
\author[Pavia]{C.~Vignoli}
\author[Princeton]{C.~Visnjic}
\author[ANL]{R.~Vondrasek}
\author[Napoli]{A.~Yushkov}

\address[Princeton]{Department of Physics, Princeton University, Princeton, NJ 08544, USA}
\address[Princeton-ChemEng]{Department of Engineering, Princeton University, Princeton, NJ 08544, USA}
\address[LNGS]{INFN, Laboratori Nazionali del Gran Sasso, Assergi (AQ) 67100, Italy}
\address[Padova]{INFN and Dipartimento di Fisica, University of Padua, Padua 35131, Italy}
\address[Felician]{Felician College, Lodi, NJ 07644, USA}
\address[AirSep]{AirSep Corporation, Buffalo, NY 14228, USA}
\address[Otis]{Kansas Refined Helium, Otis, KS 67565, USA}
\address[Harvard]{Department of Earth and Planetary Sciences, Harvard University, Cambridge, MA 02138, USA}
\address[Schlumberger]{Schlumberger Princeton Technology Center, Princeton, NJ 08550, USA}
\address[Pavia]{INFN and Dipartimento di Fisica Nucleare e Teorica, University of Pavia, Pavia 27100, Italy}
\address[Manchester]{School of Earth, Atmospheric, and Environmental Sciences, University of Machester, M13 9PL, United Kingdom}
\address[Napoli]{INFN and Dipartimento di Scienze Fisiche, University of Naples ``Federico II'', Naples 80216, Italy}
\address[LAquila]{INFN and Dipartimento di Fisica, University of L'Aquila, L'Aquila 67100, Italy}
\address[NotreDame]{Department of Physics, Notre Dame University, Notre Dame, IN 46556, USA}
\address[LindeEng]{Linde Engineering, Murray Hill, NJ 07974, USA}
\address[LindeGas]{Linde Gas, Murray Hill, NJ 07974, USA}
\address[ANL]{Argonne National Laboratories, Argonne, IL 60439, USA}
\address[Bern]{Physics Institute, University of Bern, Sidlerstrasse 5, 3012 Bern, Switzerland}
\address[Krakow]{Instytut Fizyki Jadrowej PAN, 31-342 Krakow, Poland}
\address[Houston]{Department of Geosciences, University of Houston, Houston, TX 77204, USA}
\address[Stanford]{Department of Physics, Stanford University, Stanford, CA 94305, USA}
\address[Bicocca]{INFN and Dipartimento di Fisica, University of Milano Bicocca, Milano 20126, Italy}


\begin{abstract}
We report on the first measurement of \ar\ in argon from underground natural gas reservoirs.  The gas stored in the US National Helium Reserve was found to contain a low level of \ar.  The ratio of \ar\ to stable argon was found to be \mbox{$\leq$4$\times$$10^{-17}$} (84\%~C.L.), less than 5\% the value in atmospheric argon \mbox{$( \ar/{\rm Ar}$=8$\times$$10^{-16})$}.
The total quantity of argon currently stored in the National Helium Reserve is estimated at 1000~tons.  \ar\ represents one of the most important backgrounds in argon detectors for WIMP dark matter searches.
The findings reported demonstrate the possibility of constructing large multi-ton argon detectors with low radioactivity suitable for WIMP dark matter searches.
\end{abstract}

\end{frontmatter}

\section{Introduction}
\label{sec:intro}

The existence of dark matter is well established, but its nature is unknown.  One possible candidate is a gas of Weakly Interacting Massive Particles (WIMPs) formed in the early history of the Universe.   The WIMP particle is also motivated theoretically in extensions of the standard model based on ÒsupersymmetryÓ and will be the subject of searches in upcoming experiments at the LHC at CERN.   WIMP dark matter particles,  if they exist, could be detected by observing  their collisions with ordinary nuclei as the earth moves through the gas.  Because of the low relative velocity between the target and the WIMPs, the nuclear recoils will have a small energy.   For WIMPs with a mass of $\sim$100~GeV and medium mass target nuclei, the recoil spectrum is continuous with a maximum kinetic energy of $\sim$100~keV.   The WIMP-nuclear cross section is expected to be at the weak interaction scale, and thus expected rates are small, possibly as low as a few events per ton of target per year.   Detecting WIMP dark matter could require a large detector with low background and a low threshold~\cite{dmsag}.

The noble elements---neon, argon, and xrecenon---are ideal targets for WIMPs searches as they allow detection of rare WIMP-induced nuclear recoils down to a few keV by scintillation and/or ionization. Liquid argon, in particular, is an excellent material for use as a detector of ionizing particles.  It produces copious scintillation light, allows the drift of the ionization charge over long distances, and it has been used for large detectors~\cite{amerio}.  Moreover, the difference in the stopping power between nuclear recoils and $\beta/\gamma$~events leads to a significant difference in the ratio of ionization charge to scintillation light detected and produces significant differences in scintillator pulse shapes, providing powerful tools to separate WIMP-induced events from natural radioactivity~\cite{benetti_arxiv,hitachi}.  Studies of the beta/recoil discrimination with the WARP~3.2-kg liquid argon detector demonstrate that the discrimination by pulse shape alone permits a separation of~1 recoil in~10$^8$~betas, and the beta/recoil separation by the ratio of scintillation to ionization is~1 in~10$^2$~\cite{benetti_arxiv}.

For a liquid argon detector, the separation of recoil events from $\beta$ events is particularly important because of the intrinsic background from $\beta$ decays of \ar\ , present in atmospheric argon.  The specific activity of \ar\ ($Q=565$ keV, $t_{1/2}=269$ y) is $\sim$1~Bq/kg of atmospheric argon~\cite{loosli_1983,benetti_nim}.  \ar\ is produced by cosmic ray interactions in the atmosphere, principally via the \mbox{\arforty(n, 2n)\ar} reaction~\cite{lehmann_loosli_1989,lehmann_loosli_1991}.

The WARP~3.2-kg detector~\cite{benetti_arxiv} published results from a first search for WIMPs obtaining a sensitivity comparable to the best current limits~\cite{cdms}.  The high selectivity for argon recoils should permit a sensitive WIMP search with a 140~kg liquid argon detector employing atmospheric argon, currently under construction at Laboratori Nazionali del Gran Sasso~\cite{warp-140}.  However, in spite of its favorable $\beta$/recoil discriminating power, it is highly desirable to use argon with a much lower \ar\ contamination for future, larger detectors.  Based on the proven $\beta$/recoil discrimination, a 10-fold or more reduction of \ar\ with respect to the atmospheric level would enhance the prospects of future multi-ton argon WIMP detectors.

The availability of large quantities of argon depleted in \ar\ may also enable proposed experiments to study neutrinos from reactors and from high-intensity stopped-pion neutrino sources through neutrino-nucleus elastic scattering~\cite{hagmann,scholberg}, with the potential of constraining parameters for non-standard interaction between neutrinos and matter, and of realizing precision measurements of the weak mixing angle and of the neutrino magnetic moment~\cite{scholberg}.  Thanks to the excellent properties of identification of nuclear recoils from $\beta/\gamma$ events, depleted argon could be used for the development of small portable neutrino detectors to monitor reactor sites for non-proliferation efforts~\cite{hagmann}.  Depleted argon could also be used to develop neutron detectors for port security.

Centrifugation or differential thermal diffusion are established methods for {\ar}/\arforty\ isotopic separation, but such techniques could become extremely expensive and require a long production time on a multi-ton scale.  Argon from natural gas wells is a promising source of \ar-depleted argon because \ar\ production induced by cosmic rays is strongly suppressed underground.  Shielding of target materials in deep underground reservoirs has recently played a crucial role in solar neutrino physics: the Borexino experiment measured the sub-MeV $^7$Be solar neutrinos in a liquid scintillator target~\cite{bx}, beating background from cosmic ray-induced \iso{14}C thanks to the extremely low level of the $^{14}$C/C ratio found in crude oil reservoirs ($\sim$10$^{-18}$), six orders of magnitude below typical modern carbon values~\cite{ctf}.

In the subsurface, \ar\ can be produced by a number of reactions, mainly neutron reactions on potassium, such as \iso{39}{K}(n,p)\ar~\cite{fabryka}.  Argon samples collected in groundwater from U and Th rich crystalline rocks revealed an enhanced \ar\ activity, with \ar/Ar ratios exceeding atmospheric values by up to 16 times~\cite{loosli_1989}.  The enhanced activity is likely due to an abundant local neutron flux, originating from fission or ($\alpha$,n) reactions induced by decays in the U and Th chains. The \ar/Ar ratio in subsurface gases thus depends on the local U, Th, and K concentration, on the porosity of the surrounding rocks, and may vary significantly among geological formations.

Prior to this work, measurements of the \ar/Ar ratio in natural gas wells are not available in the literature, to the best of our knowledge.\footnote{While this work was being completed, we learned of a previous unpublished attempt to measure the \ar/Ar ratio in subsurface gas~\protect\cite{loosli_1979}.}
The ubiquitous presence, in natural gas, of \iso{4}{He}---a by-product of radioactive decays in the U and Th chains---and the correlation between \arforty\ and \iso{4}{He} content~\cite{ballentine_2002} called for a measurement of the \ar/Ar ratio in underground argon to assess its potential for use in ultra-sensitive WIMP detectors.

\section{Discovery of underground argon with low level of radioactive \ar }
\label{sec:discovery}

\begin{table}[!t]
\centering
\begin{tabular}{cc}
\hline\hline
Component & Concentration by Volume \\
\hline
He & 77.3\% \\
N$_2$ & 20.3\% \\
CH$_4$ & 1.6\% \\
H$_2$ & 8000 ppm \\
Ar & 680 ppm \\
CO$_2$ &110 ppm\\
\hline\hline
\end{tabular}
\caption{Composition of crude helium extracted from the National Helium Reserve.}
\label{tab:crude_helium}
\end{table}

We report on the measurement of argon from the US National Helium Reserve, located in the Cliffside Storage Facility outside Amarillo, TX~\cite{amarillo}.  The facility stores crude helium separated by chromatography and/or cryogenic distillation from the nearby helium-rich natural gas fields.  The separation processes employed also transfer a limited fraction of the argon contained in the natural gas along with the crude helium stream.  Production takes place in nine separate plants (five in Kansas and four in Texas) so that the gas in the Reserve represents an average sampling of natural gas sites in the Texas Panhandle and southern Kansas.  All plants are connected to the US National Helium Reserve by the Helium Conservation Pipeline, run by the Bureau of Land Management.

Crude helium samples from the National Helium Reserve were collected from the Helium Conservation Pipeline at the Kansas Refined Helium plant in Otis, Kansas.  In order to minimize contamination from atmospheric argon in the samples, the booster pump used to transfer crude helium into the bottles was driven by a separate stream of crude helium.  The composition of the crude helium, as measured by mass spectrometry, is given in Table~\ref{tab:crude_helium}.

The crude helium was processed with a novel Pressure Swing Adsorption (PSA; see Ref.~\cite{knaebel_hill}) plant developed at Princeton.  The unit is capable of processing the gas stream to remove the strongly adsorbing components (N$_2$, CO$_2$, H$_2$S, CH$_4$, and heavier hydrocarbons), concentrating Ar, He, and H.  The concentration of strongly adsorbed gases---N$_2$, CH$_4$, and CO$_2$---was reduced by a factor $>$10$^4$~\cite{loer}.  At the output of the PSA plant, argon was trapped at 76~K on liquid N$_2$-cooled activated charcoal and separated from helium and hydrogen.  The gas de-sorbed from the charcoal trap was $\sim$80\% argon, the remaining 20\% being mainly helium (with traces of $N_2$ and hydrogen).  Contamination from atmospheric argon is negligible, the whole gas separation system being leak tight to 10$^{-9}$~mbar$\cdot$l/s.

The PSA plant used for this work consists of two columns filled with zeolites 13X.  Zeolites have a good selectivity for Ar versus N$_2$: the adsorption capacity for Ar at 300~K and atmospheric pressure is 2~liter/kg, compared to an adsorption of O$_2$ of 9~liter/kg~\cite{sebastian}.  The quantity of gas adsorbed depends linearly on the partial pressure for total pressures up to 100~kPa.  CO$_2$, H$_2$S, CH$_4$, and heavier hydrocarbons are much more strongly adsorbed than N$_2$ on zeolites 13X.  H$_2$ and He are adsorbed less strongly than Ar, and O$_2$ behaves in a very similar way to Ar.  Our PSA plant exploit this difference in adsorption to separate the individual components into a product and a waste stream.  Concentration of components with low adsorption is strongly enriched in the product stream.  By alternating the pressure and direction of the gas flow, and running the two columns in opposite phases through consecutive high pressure feed cycles and low pressure purge cycles, the separation of the two streams can be performed in a continuous cycle.  With this configuration, the adsorbent is self-regenerated during the low pressure purge cycle.

The gas flow in our PSA plant is directed by solenoid valves.  The control system for the solenoid valves consists of a set of relays interfaced with a programmable automation controller, designed for industrial control.  The software for the control system was developed in Princeton using the {\tt LabVIEW FPGA} platform.

\begin{center}
\begin{figure}
\includegraphics[width=0.5\textwidth]{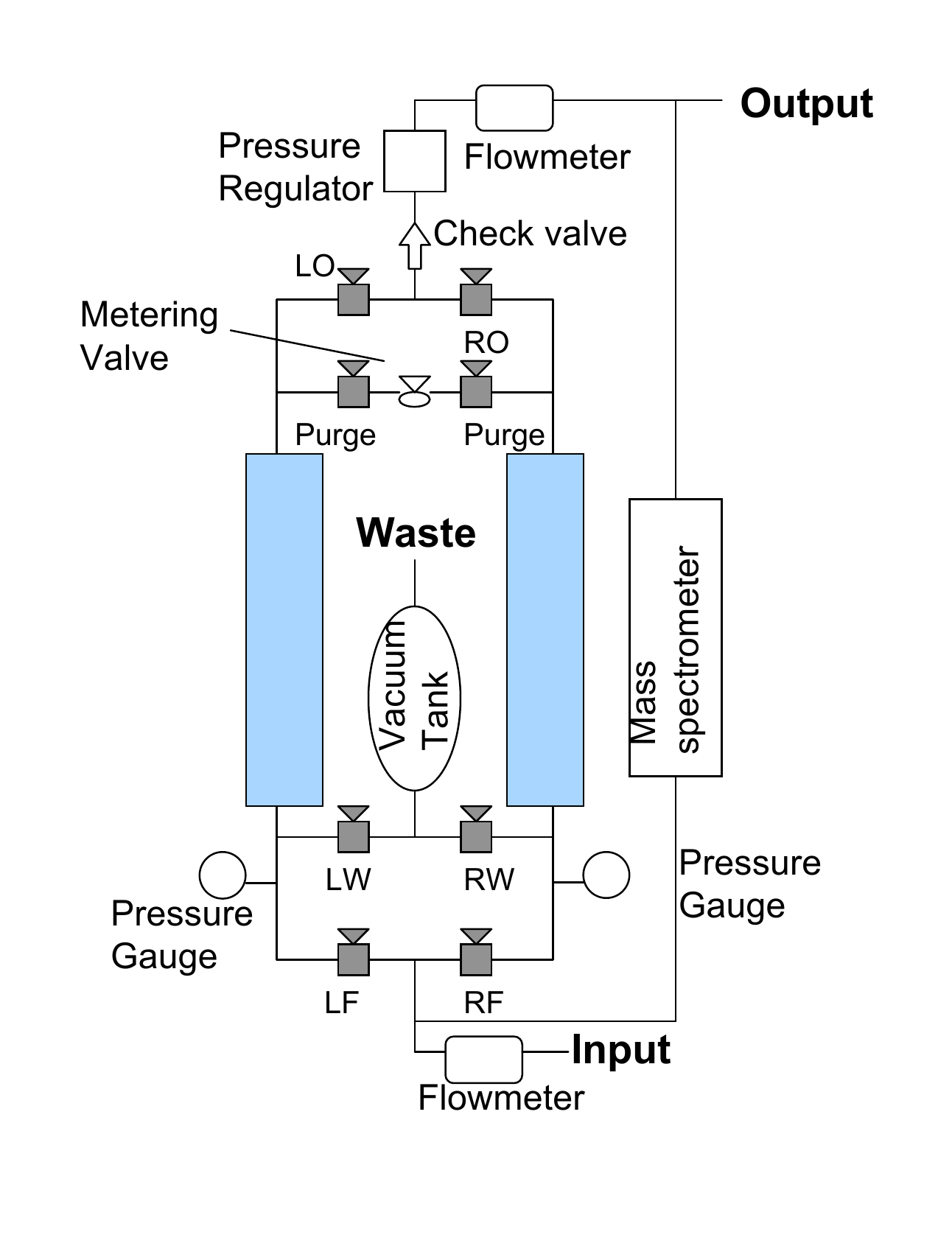}
\caption{Diagram of the PSA plant developed at Princeton.  See text for details.}
\label{fig:psa}
\end{figure}
\end{center}

Measuring \ar\ at or below atmospheric concentration is very challenging.  A review of the methods used to measure trace levels of rare gas isotopes is given in reference~\cite{loosli_purtschert}.  The standard method, used in this work, is direct counting in underground, low-background proportional counters~\cite{loosli_1980,collon}.  \ar\ concentrations detectable by this method are typically $\sim$5\% of atmospheric value~\cite{loosli_purtschert}.  We note that activation of \iso{40}Ar in the sample during the three months exposure to the surface flux of cosmic rays is negligible within the accuracy of our measurement.
\footnote{The activation on the surface is lower than the average activation time in the atmosphere, given that most of the 39Ar production takes place in the upper layers of the atmosphere.
The e-folding activation time in the atmosphere is the 39Ar meanlife, i.e. 269 years.
Therefore an upper limit for the activation in three months is given by: 3 mos/269 yr = 0.1\%.}

The measurements reported here were performed at the Low Level Counting Underground Laboratory at the Physics Institute at the University of Bern, located underground at a depth of 70~m water equivalent.  This is, to our knowledge, the only facility where a number of low background proportional counters are dedicated to routine measurements of the \ar/Ar ratio in samples of various origins for environmental and climatic studies~\cite{schlosser}.

\begin{center}
\begin{figure}[!t]
\includegraphics[width=0.5\textwidth]{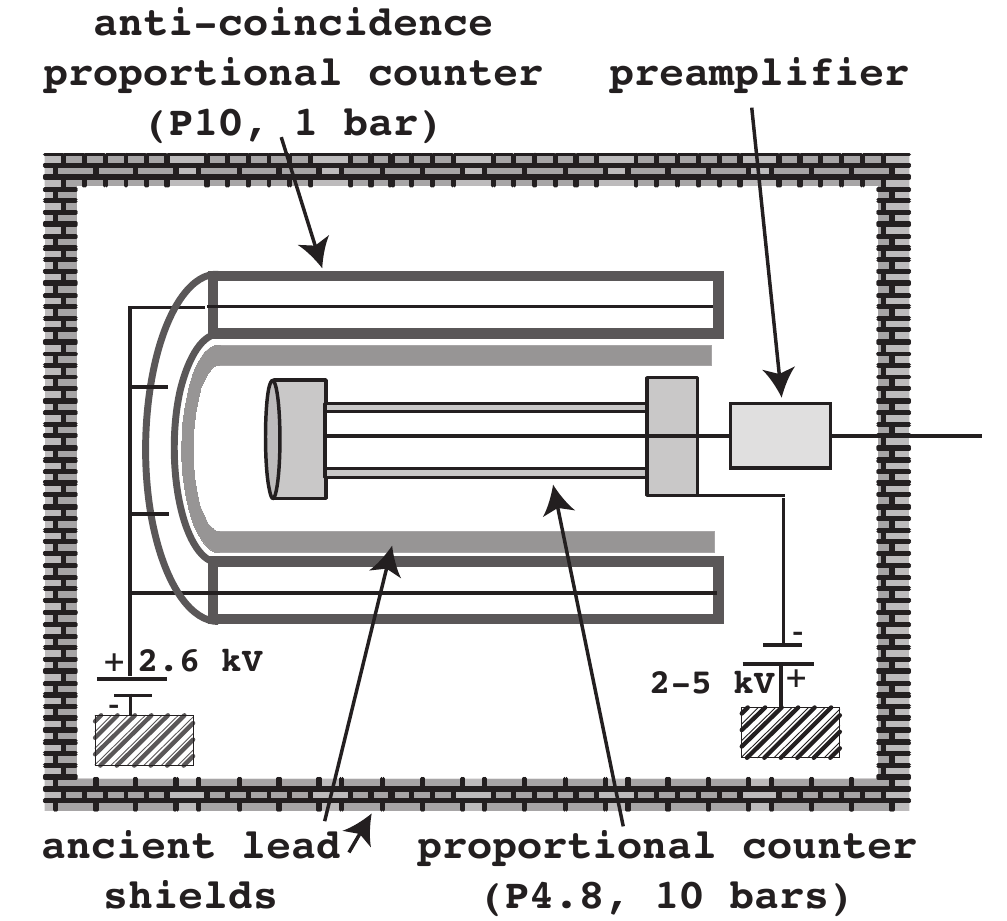}
\caption{Schematic of the setup utilized to measure the \ar/Ar ratio at the University of Bern Low Level U. Lab..  For \ar\ measurements, the counters are filled with a P4.8 mixture (4.8\%~methane, 95.2\%~argon) at pressures in the range 5--25~bars. The measurements are performed at a depth of 35~meters. Low activity lead shield and an anti-coincidence arrangement provide a further reduction of the background count rate.}
 \label{fig:propcounter}
\end{figure}
\end{center}

The laboratory is located at a depth of 35~m, providing a reduction of the muon flux by a factor of $\sim$10~\cite{loosli_1980}.  The lab walls are constructed utilizing a special concrete selected for its low radioactivity content, in order to minimize the gamma-ray flux within the lab.  The main components of the detector and of the counting system are shown in Fig.~\ref{fig:propcounter}.  A 100~cm$^3$ proportional counter (dimensions: 25~cm length and 2.2~cm diameter), built of high-conductivity oxygen-free copper, is filled with the sample gas at a pressure of 10~bars and placed in a cylindrical lead shield 5~cm thick (see Fig.~\ref{fig:propcounter}).  Only a fraction of the $\beta$-particle decay energy is released within the counter before they reach its wall.  The deposited energy is recorded by a 7-bit Multi-Channel Analyzer (MCA), with a linear energy range of 0--35~keV; events with energies greater than~35~keV are recorded as saturated events in the last MCA channels and included in the analysis.

For background reduction, the proportional counter is constructed with low background materials and uses an anti-coincidence proportional counter to reduce background.  The assembly is inserted in a second, larger, cylindrical proportional detector, which acts as an anti-coincidence counter.  The passive shielding is complemented by an external lead shield 12~cm thick.  The lead shields were built using lead from ancient ship wrecks with very low \iso{210}Pb~intrinsic activity.

The detector background is comparable to the rate due to \ar\ in atmospheric argon and must be determined accurately.  The background of the whole counting system is measured in a separate run using argon, depleted in \ar\ by at least a factor of~20-50 through differential thermal diffusion at Monsanto Co. (Miamisburg OH, USA; see also Ref.~\cite{loosli_1983}).  The \ar/Ar content  of such sample was confirmed to be, at 99\% C.L., lower than 3\% of the \ar/Ar ratio in the atmosphere by comparison with gas samples extracted from ancient ice---independently dated at more than 1000~old---and with gas samples from several aged groundwater sources~\cite{loosli_1983,lehmann_2003,purtschert}.  

Three measurements are performed to measure the \ar\ content in the sample of interest:  one with the \ar-depleted gas to measure intrinsic background, one with standard atmospheric argon for reference, and one with the underground sample itself.  The \ar\ activity of the sample under investigation is evaluated by subtracting the detector background from the total measured activity and is then compared to the \ar\ activity in atmospheric argon (background subtracted).  Results are expressed in terms of the \ar/Ar ratio relative to the atmospheric ratio.

The energy spectrum above threshold was analyzed with the routine procedure used for all \ar\ measurements performed in the Low Level Counting Underground Laboratory (for a detailed review of the procedure, see Ref.~\cite{loosli_1983}).  Table~\ref{tab:ROI} reports the count rate from the three samples in four regions of interest (ROI).  The count rates in every ROI for the argon sample from the US National Helium Reserve are in good agreement with rates from the reference \ar-depleted gas.  The best estimate for the \ar\ activity is obtained from the total count rate in the whole spectrum (Channels~20--127) and indicates a \ar/Ar ratio \mbox{$\leq$4$\times$10$^{-17}$}, which is less than~5\% of the \ar/Ar ratio in atmospheric argon.  The depletion factor from the atmospheric activity of the sample is therefore $\geq$20 at 84.1\% C.L. ($\geq$10 at the 97.7\% C.L.).

\begin{table*}[t!]
\centering
\begin{tabular}{lcccc}
\hline\hline
\multicolumn{5}{c}{Argon Sample from}  \\
\multicolumn{5}{c}{National Helium Reserve, Amarillo, TX}  \\
\hline
 &\multicolumn{4}{c}{Count Rate [$\mu$Bq]}  \\
 &Ch.~20--50 &Ch.~51--102 &Ch.~103--127 &Ch.~20--127 \\
\hline
1) Underground Ar
	&460$\pm$21	&480$\pm$21		&1096$\pm$32		&2036$\pm43$	 \\
2) \ar-depl. Reference
	&454$\pm$23	&512$\pm$25		&1068$\pm$35		&2035$\pm$49	 \\
3) Atmospheric Ar
	&542$\pm$30	&855$\pm$37		&2228$\pm$60		&3625$\pm$77 \\
4) (Under. Ar) - (Ref.)
	&6$\pm$30	&-33$\pm$32		&28$\pm$48		&1$\pm$65 \\
5) (Atm. Ar) - (Ref.)
	&88$\pm$38	&342$\pm$45		&1160$\pm$70		&1589$\pm$91 \\
\hline
\multicolumn{4}{l}{6) [(Under. Ar) - (Ref.)]/[(Atm. Ar) - (Ref.)]} &0.00$\pm$0.04 \\
\multicolumn{4}{l}{7) (\ar/Ar)$_{\rm und}$/(\ar/Ar)$_{\rm atm}$} &$<$0.05 \\
\hline\hline
\end{tabular}
\caption{Count rate for different ROIs: Channels 20--50, 51--102, 103--127, and the whole spectrum above threshold, 20--127.  The rates are reported for the three gas samples under consideration (lines~1--3: argon from the US National Helium Reserve, \ar-depleted reference, atmospheric argon).  The majority of the counts are due to the detector background.  This background is measured by using a \ar-depleted reference gas and subtracted from the activities measured for the sample from the US National Helium Reserve (line 4) and for atmospheric argon (line 5).  The \ar/Ar ratio for the underground sample is 0.00$\pm$0.04 relative to the atmospheric value, where the uncertainty comes from the statistical errors associated with the three measurements (line 6).  The best estimate becomes 0.00$\pm$0.05 when taking into account the uncertainty on the \ar/Ar ratio for the reference sample, which is combined in quadrature with the statistical uncertainty (line 7).}
\label{tab:ROI}
\end{table*}

\section{Conclusions}
\label{sec:conclusions}

This result is, to the best of our knowledge, the first measurement of the concentration of \ar\ in subsurface gas.  It represents an upper limit which is based on the background of the proportional counting system. It therefore motivates the development of new and potentially more sensitive techniques, such as Accelerator Mass Spectrometry (AMS)~\cite{collon} and \ar-decay counting in liquid argon detectors~\cite{benetti_nim} for a more accurate determination of the \ar/Ar ratio.  AMS measurements at the ATLAS facility at Argonne National Laboratory are already capable of measuring \ar/Ar ratios around 5\% of the atmospheric levels, comparable to the best limits achievable with gas proportional counters.   With improvements currently under development to reduce the \iso{39}{K} background in the detector while still allowing a high beam current, a new limit of approximately 10 times lower is expected.  \ar-decay counting in the 3.2-kg WARP detector already reached an accuracy of 10\% of atmospheric levels, and a similar, specially designed, low background, $\sim$10-kg detector could achieve a sensitivity 100 times higher, down to 1 part in 10$^3$ of the atmospheric activity.

The discovery of low \ar/Ar ratio in the US National Helium Reserve is part of a larger program of exploration of several possible underground argon sources.  During this investigation, the authors developed the technology required to separate and collect large quantities of argon from natural gas wells.  The quantity of argon processed by the Kansas Refined Helium plant is 25 tons per year.  Production of a 10 ton target of liquid argon for a dark matter search experiment would take a 1-yr production campaign after construction and commissioning of dedicated separation plants.

These findings lead the way for future multi-ton, low background, argon detectors for WIMP dark matter, able to reach sensitivities for the WIMP-nucleon cross section of~10$^{-46}$~cm$^2$ or smaller.  The availability of large quantities of argon depleted from \ar\ will also be beneficial for studies of neutrino properties through neutrino-nucleus coherent scattering.  It may also enable the construction of small and portable neutrino detectors for reactor monitoring in non-proliferation efforts and of neutron detectors for port security.

\section{Acknowledgments}
\label{sec:ack}

This work was supported by the US National Science Foundation under Grant no.\,0704220.  This work was supported in part by the US Department of Energy, Nuclear Physics Division, under contract DE-AC02-06CH11357.  The WARP program is funded by the italian Instituto Nazionale di Fisica Nucleare and by the US National Science Foundation.  A.\,M.~Szelc has been in part supported by a grant of the President of the Polish Academy of Sciences and by the MNiSW grant 1P03B04130.  The authors express their gratitude to the Kansas Refined Helium Co., a division of Linde, for access to the Helium Conservation Pipeline and assistance with sampling and transport.  The authors thank W.~Brinkman, D.~Marlow, J.~Russell, and P.~Wraight  for their support and encouragement.

\end{document}